\documentclass[apjl]{emulateapj}
\usepackage{color}

\newcommand{\OIII}{\mbox{[O\,\textsc{iii}]}}

\newcommand{\FeII}{\mbox{Fe\,\textsc{ii}}}
\newcommand{\kms}{km s$^{-1}$}
\newcommand{\Ha}{H$\alpha$}   
\newcommand{\Hb}{H$\beta$}  
\newcommand{\msigma}{M$_{\rm BH}$-$\sigma_*$}
\newcommand{\msun}{M$_{\odot}$}
\newcommand{\VOIII}{V$_{\rm OIII}$}
\newcommand{\SOIII}{$\sigma_{\rm OIII}$}
\newcommand{\LOIII}{L$_{\rm OIII}$}
\newcommand{\SVD}{$\sigma_{*}$}

\newcommand{\ergs}{erg s$^{-1}$}

\shorttitle{The correlation of outflow kinematics with star formation rate}
\shortauthors{Woo et al.}

\begin{document}

\title{The correlation of outflow kinematics with star formation rate - Gas outflows in AGNs. VI.}
\author{Jong-Hak Woo$^{1}$}
\author{Donghoon Son$^{1}$}
\author{Suvendu Rakshit$^{1,2}$}

\affil{$^{1}$Astronomy Program, Department of Physics and Astronomy, Seoul National University, Seoul 08826, Republic of Korea; woo@astro.snu.ac.kr}
\affil{$^{2}$Finnish Centre for Astronomy with ESO (FINCA), FI-20014 University of Turku, Finland}

\begin{abstract}

We investigate the connection between ionized gas outflows and star formation activity using a large sample of type 1 and 2 AGNs
with far-IR detections or star formation rate (SFR) estimates. 
The strength of ionized gas outflows, measured by the velocity dispersion and velocity shift of the \OIII\ emission line, clearly shows a correlation with
SFR. The connection between specific star formation rate (sSFR) and \OIII\ gas velocity dispersion 
indicates that AGNs with stronger outflows are hosted by galaxies with higher SFR. Compared to star-forming galaxies in the main sequence, both type 1 and type 2 AGNs show sSFR similar to that of non-AGN galaxies, indicating no instantaneous AGN feedback, while sSFR is higher (lower) for AGNs with stronger (weaker) outflows than that of main sequence galaxies. These results are consistent with a delayed AGN feedback scenario. However, it is also possible that a decease/increase of gas fraction may cause the correlation without AGN feedback. 

\end{abstract}

\keywords{galaxies: active --- galaxies: kinematics and dynamics}

\section{Introduction}

The role of supermassive black holes in galaxy evolution has been a crucial subject over the last two decades 
since the scaling relations between black hole mass and host galaxy properties had been reported 
\citep[e.g.,][]{fm+00,ge+00}. While cosmological galaxy simulations often adopt active galactic nuclei (AGN) feedback models \citep[e.g.,][]{Dimatteo+05,Croton06,Dubois+13,deGraf+15}, the nature of black hole - galaxy coevolution is yet to be fully understood. 

Gas outflows may affect star formation in galaxy scales by delivering the energy output from AGN. Various observations confirmed large scale AGN-driven gas outflows \citep[e.g.,][]{Nesvadba+06, Liu+13, Harrison+14, Husemann+14, Husemann+16}, suggesting that gas outflows are an effective channel of AGN feedback. Two opposite effects are expected \citep{2020A&ARv..28....2V}. First, AGN feedback suppresses SF by removing gas
from host galaxies or by preventing gas cooling \citep{Silk+98, Fabian12, 2012NewAR..56...93A}. Second, outflows may compress ISM, enhancing SF \citep{Zubovas+13, Ishibashi+14}.
Various observational studies of individual galaxies reported signatures of both negative and positive feedback, revealing that the nature of AGN feedback is complex \citep[e.g.,][]{Cresci+15, Carniani+16, Vilar-Martin+16, Karouzos+16b, Shin+19}. Albeit with decade-long efforts to unveil the role of AGN feedback, smoking gun evidence of AGN feedback suppressing star formation is still missing \citep{Harrison+18}.

It is well known that AGN luminosity broadly correlates with SF luminosity. For example, AGN bolometric luminosity traced by
X-ray or optical continuum shows good correlation with SF luminosity measured from various indicators \citep[e.g.,][]{Netzer09,Diamond-Stanic+12,Woo+12,Matsuoka15}
while at high redshifts, the connection between the two may be different \citep[e.g.,][]{Rosario+12,Santini+12, 2019MNRAS.486.4320R, 2020ApJ...888...78S}. 
These results indicate that on average AGN and SF activities are simultaneously on-going, and gas supply feeds both AGN and SF.

As there is a broad range of AGN energetics, as demonstrated by the Eddington ratio distribution \citep[e.g.,][]{Woo+02, 2006ApJ...648..128K, Brandon+13}, it is reasonable to expect a systematic difference of AGN feedback between more energetic or stronger outflow AGNs and less energetic or weaker outflow AGNs.
In fact, an increasing trend of specific star formation rate (sSFR) with increasing Eddington ratio has been reported based on low-z AGNs \citep{Shimizu+15, Ellison+16b, Woo+17}. The link between sSFR and AGN Eddington ratio is likely due to AGN outflows. For example, \citet{Woo+17} reported that the sSFR systematically decreases with decreasing AGN outflow strength using a large sample of type 2 AGNs at z $<$ 0.3. 

To understand AGN feedback via gas outflows, we have been performing a series of studies, focusing on ionized gas outflows in low-z AGNs. 
For a statistical investigation of gas outflows we use a large sample of type 1 and type 2 AGNs from Sloan Digital Sky Survey (SDSS). In the first of this series, \citet{Woo+16} reported a detailed study of the demography of ionized gas outflows using the \OIII$\lambda$5007 kinematics
\citep[for \Ha\ kinematics, see][]{Kang+18}, showing that more than 90\%\ of high-luminosity AGNs has gas outflows. 
In the fifth of this series, \citet{Rakshit+18} used a large sample of type 1 AGNs at similar redshifts, reporting that $\sim$90\%\ of type 1 AGNs shows
outflow signatures in the \OIII\ line profile, demonstrating the prevalence of outflows in AGNs.
Based on the statistical study of gas outflows, we investigated how SFR is affected by AGN outflows by comparing SFR with outflow strength
using type 2 AGNs \citep{Woo+17}. The main results indicate that strong outflow AGNs show similar sSFRs compared to main sequence star-forming galaxies (SFG), while the sSFR of no outflow AGNs is much lower than that of main sequence SFGs.  In other words, although AGNs show strong outflows, SF activity is not suppressed, suggesting no instantaneous effect of AGN feedback or delayed influence, presumably due to the fact that dynamical time is required for outflows to impact on star formation over large galaxy disk scales.

The main limitation of our previous study is that black hole mass and bolometric luminosity are estimated with large uncertainties, and the selected type 2 AGN
sample based on the emission line flux ratios may be biased against low luminosity AGNs. To compensate these obstacles, we combine type 1 and type 2  AGN samples together. While each sample has pros and cons, the combined sample can provide better understanding as the two samples are complementary. In this paper, we present the correlation between star formation and AGN outflow strength using a large sample of type 1 and type 2 AGNs, respectively collected from \citet{Woo+17} and \citet{Rakshit+18}. 
Throughout the paper, we use the cosmological parameters as
$H_0 = 70$~km~s$^{-1}$~Mpc$^{-1}$, $\Omega_{\rm m} = 0.30$, and $\Omega_{\Lambda} = 0.70$.

\section{Sample \& Analysis}

\subsection{Sample selection}

We constructed a large sample of type 1 and type 2 AGNs. First, we chose a sample of type 1 AGNs at z$<$0.3, which was generated by \citet{Rakshit+18} to investigate ionized gas outflows using \OIII. This sample is composed of 5,189 objects with well determined \OIII\ velocity shift and velocity dispersion. The detailed analysis of outflows in comparison with other AGN properties is presented by \citet{Rakshit+18}. 
To increase the dynamic range of type 1 AGNs, we adopted a sample of 609 lower luminosity AGNs, which were identified as type 1 AGNs based on the presence of a broad component in the \Ha\ emission line profile by \citet{Woo+14, Eun+17} using the SDSS type 2 AGN catalogue of \citet{Bae&Woo16, Woo+16}. 
Second, we added a large sample of type 2 AGNs using the catalogue generated by \citet{Woo+16}. For these type 2 AGNs at the matched redshift z$<$0.3, \citet{Woo+17} presented a detailed study of ionized gas outflows and SF.  
Among the AGNs in the combined sample, we obtained SFR for 900 type 1 AGNs and 20,526 type 2 AGNs (see \S2.3). Thus, we investigate the impact of AGN gas outflows on SF activity, using the final sample of 21,426 AGNs.

\subsection{\OIII\ kinematics}

The details of the kinematical properties of the ionized gas of the sample were presented by \citet{Woo+16} and \citet{Rakshit+18} for type 2 and type 1 AGNs, respectively. Here we summarize the measurement procedure for completeness. The \OIII\ line at 5007\AA\ was used to trace gas outflows in the ionized phase. 
We measured the velocity (1st moment) and velocity dispersion (2nd moment) of the total line profile using the best-fit model of \OIII. With the measured velocity of \OIII\ we calculated velocity shift with respect to systemic velocity that was
determined based on either stellar absorption lines, or the peak of the narrow component of the \Hb\ line if stellar lines were not detected. 
Note that the peak of \Hb\ is close to systemic velocity albeit with small additional uncertainties as we showed the direct comparison between 
the \Hb\ peak velocity and stellar-based systemic velocity in our previous study \citep[see Figure 2 of][]{Rakshit+18}. 

The 1st moment of \OIII\ is systematically different from the peak of \OIII, if the line profile is fitted with a double Gaussian model, as the wing component significantly contributes to the flux-weighted line profile. Since the narrow core component of \OIII\ typically represents the gravitational potential of host galaxies, various studies used only the wing component to calculate the velocity shift and velocity dispersion of \OIII\ gas. 
In our study, however, we used the total line profile since the core and wing components often show a similar line width, 
leading to a difficulty of separating the outflow component from the gravitational component. For comparison, we additionally presented the measurements of the kinematics based on the wing component only (see Section 3).

To fit the \OIII\ line profile, we first subtracted the continuum, which was fitted with stellar population models, AGN continuum and \FeII\ emission complex. 
Then, we fitted the \OIII\ line with a double Gaussian model.
If  the \OIII\ line profile showed a prominent wing component, i.e., the amplitude-to-noise (A/N) ratio of the 2nd Gaussian component is larger than 3, we accepted the 2nd Gaussian component. Otherwise, we fitted \OIII\ with a single Gaussian model. 
Using the best-fit model, we then calculated the 1st and 2nd moments of
the line profile as 
\begin{eqnarray}
\lambda_{0} = {\int \lambda f_\lambda d\lambda \over \int f_\lambda d\lambda}.
\end{eqnarray}
\begin{eqnarray}
[\Delta\lambda_{\OIII}]^2  = {\int \lambda^2 f_\lambda d\lambda \over \int f_\lambda d\lambda} - \lambda_0^2, 
\end{eqnarray}
where $f_\lambda$ is the flux at each wavelength.
After that, we determined the velocity shift (\VOIII) of \OIII\ with respect to the systemic velocity, and the velocity dispersion (\SOIII) from the second moment. 
The uncertainties of the measurements were constrained from Monte Carlo simulations, which generated 100 mock spectra by randomizing flux with flux error, and provided 100 measurements based on the best-fit model of each spectrum.
We adopted the 1$\sigma$ dispersion of the distribution as the measurement uncertainty.

\subsection{IR-based SFR}

The star formation rate (SFR) of AGN host galaxies is difficult to directly measure due to the contamination from AGN emission to SFR indicators \citep[e.g.,][]{Matsuoka15, 2016MNRAS.457.2703R}. For this study, we used IR luminosity estimates or direct observations to calculate SFR.    
First, we selected FIR-detected AGNs by utilizing the AKARI/Far-infrared Surveyor (FIS) all-sky survey
bright source catalog (Yamamura et al. 2010). The catalogue provided flux measurements in four different bands, namely, 
65, 90, 140, and 160 $\mu$m. Among them we adopted high quality detection (i.e., quality flag FQUAL = 3) in the 90 $\mu$m
band, which is the closest to the peak of the SED in the IR spectral range.  
We cross-matched our AGN sample with the 90 $\mu$m catalogue with a matching radius of 18\arcsec, which
represents the 3$\sigma$ position error in the cross-scan direction of the FIS. In this process,
we obtained a 90 $\mu$m counterpart, respectively, for 53 type 1 and 396 type 2 AGNs. 
Since the AKARI/FIS all sky survey in the 90 $\mu$m band was shallow with a 5$\sigma$-detection limit of 0.55 Jy,
only a small fraction of
our AGNs were detected. 
In addition, we used the Herschel/PACS (Photodetector Array Camera and Spectrometer) Point Source Catalogue (Pilbratt et al. 2010), which 
provided flux measurements in the 70 $\mu$m,  100 $\mu$m, and 160  $\mu$m bands\footnote{DOI: https://doi.org/10.5270/esa-rw7rbo7}.
Using a matching radius of 3$''$  with a criterion of S/N $>$ 3, 
we obtained a counterpart in the 100 $\mu$m band, respectively, for 103 type 1 and 277 type 2 AGNs. 

For these FIR-detected AGNs, we calculate the SFR based on the Equation by \citet{Kennicutt+98}:
\begin{equation}
\rm SFR_{FIR} (M_{\odot} yr^{-1}) = 4.5 \times 10^{-44}~ L_{FIR} (erg~s^{-1})
\end{equation}
The AKARI 90 $\mu$m and Herschel 100 $\mu$m fluxes represent a slightly different part of the SED, and the redshift of each object changes 
the rest-frame spectral range. We ignored these effects for simplicity since these effects are only 1-2\% level \citep[see][]{Matsuoka15}. Note that while L$_{\rm FIR}$ in Eq. 3 represents the total luminosity integrated over the near-, mid-, and far-IR, we used the monochromatic luminosity 
at 90 or 100 $\mu$m from AKARI/FIS and Hershel/PACS. Thus, the derived SFR for each object has considerable uncertainties. For our statistical comparison, these effects are relatively small since the FIR luminosity dominates over the entire IR range.

As the sample size of the AGNs with direct FIR detections is still small, we increased the sample by adopting less reliable SFR estimates. 
As we performed in our previous study with type 2 AGNs \citep{Woo+17}, we utilized the catalogue of \citet{Ellison+16a}, which provided the total IR luminosity estimates for $\sim$330,000 SDSS galaxies based on the artificial neural network (ANN) technique, which were tested and trained based on the sample of 1136 galaxies in the Herschel-SDSS Strip 82. We adopted the SFR converted from the total IR luminosity by  \citet{Ellison+16a}, 
who used a modified equation for a Chabrier initial mass function:
\begin{equation}
\rm log~ SFR (M_{\odot} yr^{-1}) = log~ L_{\rm IR} (\rm erg~ s^{-1}) - 43.59
\end{equation}
Among available SFR estimates, we limited the sample using the uncertainty of ANN-predicted L$_{\rm IR}$, $\sigma_{\rm ANN}$ $<$ 0.3 \citep[see][]{Ellison+16a}. Based on the cross-matching, we obtained SFR for 835 objects out of 4,538 type 1 AGNs. In the case of type 2 AGNs, we collected SFR for 19,864 targets out of 22,324 objects in our previous study \citep{Woo+17}. 
If we used more reliable SFR estimates with the best-quality criterion (i.e., $\sigma_{\rm ANN}$ $<$ 0.1), we obtained 521 type 1 AGNs and
12,124 type 2 AGNs. We will use these SFR estimates for statistical comparison between outflows and SF activity. 

For a consistency check, we compare FIR-based SFR with the SFR estimate from \citet{Ellison+16a}, using a subsample of 428 AGNs 
in Figure 1. We find that while the scatter is relatively large (0.41 dex), the correlation is reasonably acceptable with the best-fit slope of 1.14$\pm$0.01, which is close to unity. If we only use the best-quality SFR estimates with $\sigma_{ANN}$ $<$ 0.1, we obtain a similar slope (1.17$\pm$0.01) and scatter (0.42 dex). 
The SFR based on FIR luminosity is on average 0.06 dex larger than that adopted from \citet{Ellison+16a}. Note that for a given total IR luminosity, Eq. 3 provides a 0.24 dex smaller SFR than Eq. 4. On the other hand, 
the monochromatic FIR luminosity at 90 or 100 $\mu$m may overrepresent the total IR luminosity, leading to overestimation of SFR. Overall, the combined effects results in a small systematic difference between the two. Thus, we will use either FIR-based SFR or ANN-based SFR without further calibration.

\subsection{Sample properties}

Since the sample of AGNs with available SFR is a subset of the total SDSS AGN sample, 
we investigate whether the sample provides a large enough dynamic range compared to the SDSS
AGNs. By comparing the distribution of stellar mass and \OIII\ luminosity (\LOIII), we check whether the sample is biased toward low (or high) AGN luminosity. 
Note that stellar mass is taken from the MPA--JHU value-added catalog\footnote{http://www.mpa-garching.mpg.de/SDSS/}.
In Figure 2 we present the stellar mass and the \OIII\ luminosity distribution of the total AGN sample, which was studied for ionized gas outflows (top panel), and the final sample of AGNs with available SFR for this study (bottom panel). 
The stellar mass of the total sample ranges from 10$^{9.5}$ to 10$^{11.5}$ \msun\ and \LOIII\ covers from 10$^{38}$ to 10$^{43}$ \ergs. 
While the stellar mass distribution of the selected AGNs for this study is consistent with that of the total sample, the mean \LOIII\ is lower than that of the total sample, particularly for type 1 AGNs. 
We do not claim the subsample is complete. Nevertheless, we demonstrate that this subsample
covers a large range of AGN luminosity for studying the connection between AGN energetics and SF activity. 

In addition to the host galaxy properties, we also determine the properties of black holes. 
For black hole mass estimates, we use the scaling relations, e.g., black hole mass-stellar velocity dispersion (\msigma) relation \citep{woo+15,Park+15},
and black hole mass - stellar mass relation \citep{marconi&hunt03} 
for type 2 AGNs. As we pointed out in Paper I, the range of black hole mass significantly changes depending on the adopted scaling relation. We adopted the black hole mass estimates based on the black hole mass - stellar mass relation since the obtained black hole mass range is more reasonable \citep[see, for details,][]{Woo+17}. 
In the case of type 1 AGNs, we measured the line width of the broad \Hb\ line (or \Ha, if not available) and AGN monochromatic luminosity 
at 5100\AA\ (or \Ha\ luminosity, if not available), to calculate black hole mass using the single-epoch mass estimator calibrated by \citet{woo+15}. 
Note that the choice of the black hole mass estimates do not change the main results presented in this paper. 

\begin{figure} 
\centering
\includegraphics[width=0.45\textwidth]{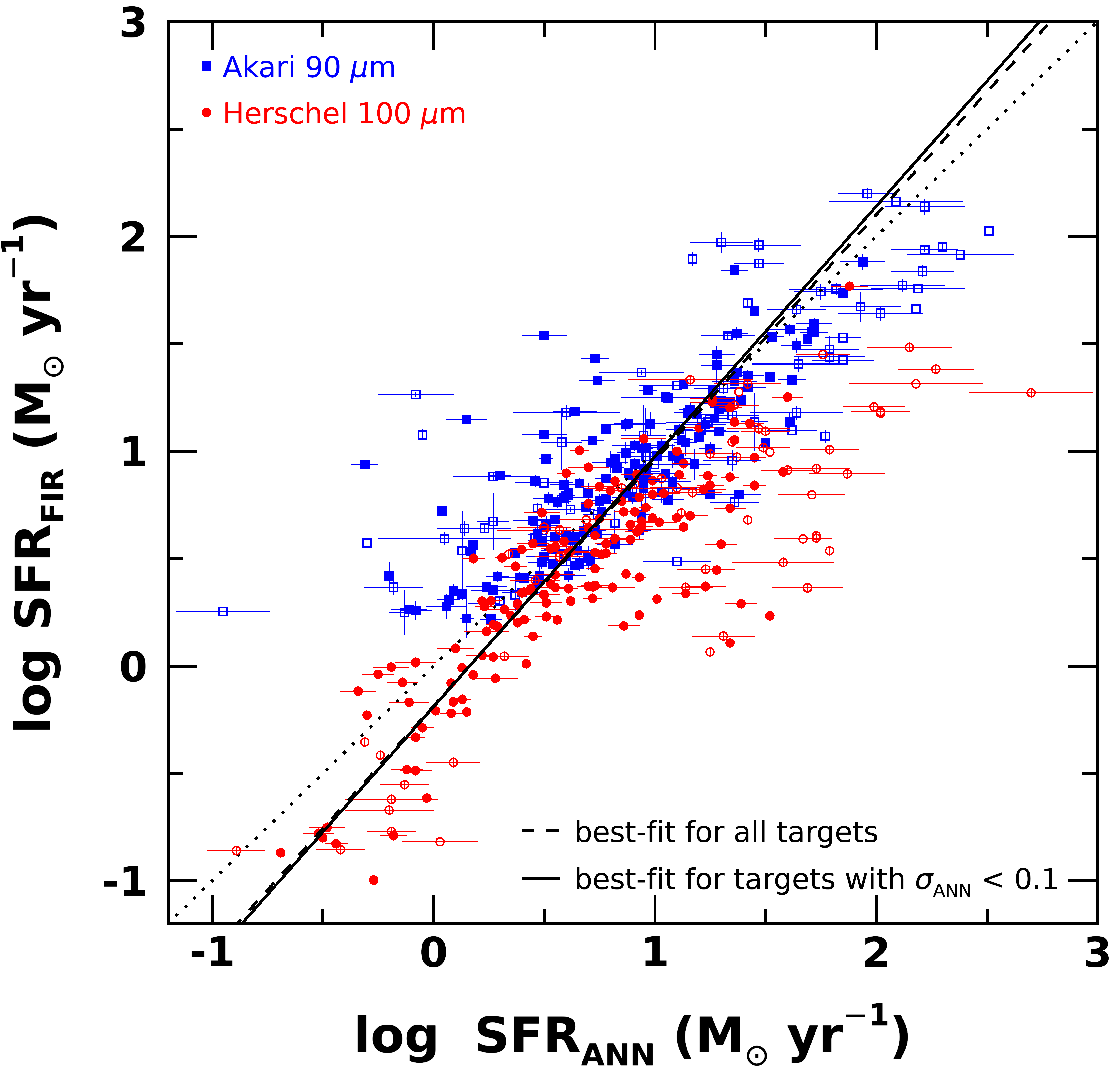}
\caption{
Comparison of the FIR detection based SFR with the SFR estimates based on the ANN analysis of \citet{Ellison+16a}.
The best SFR estimates with $\sigma_{\rm ANN}$ $<$0.1 (filled symbols) as well as good SFR estimates with 
$\sigma_{\rm ANN}$ $<$0.3 are included.
}
\end{figure}

\begin{figure} 
\centering
\includegraphics[width=0.42\textwidth]{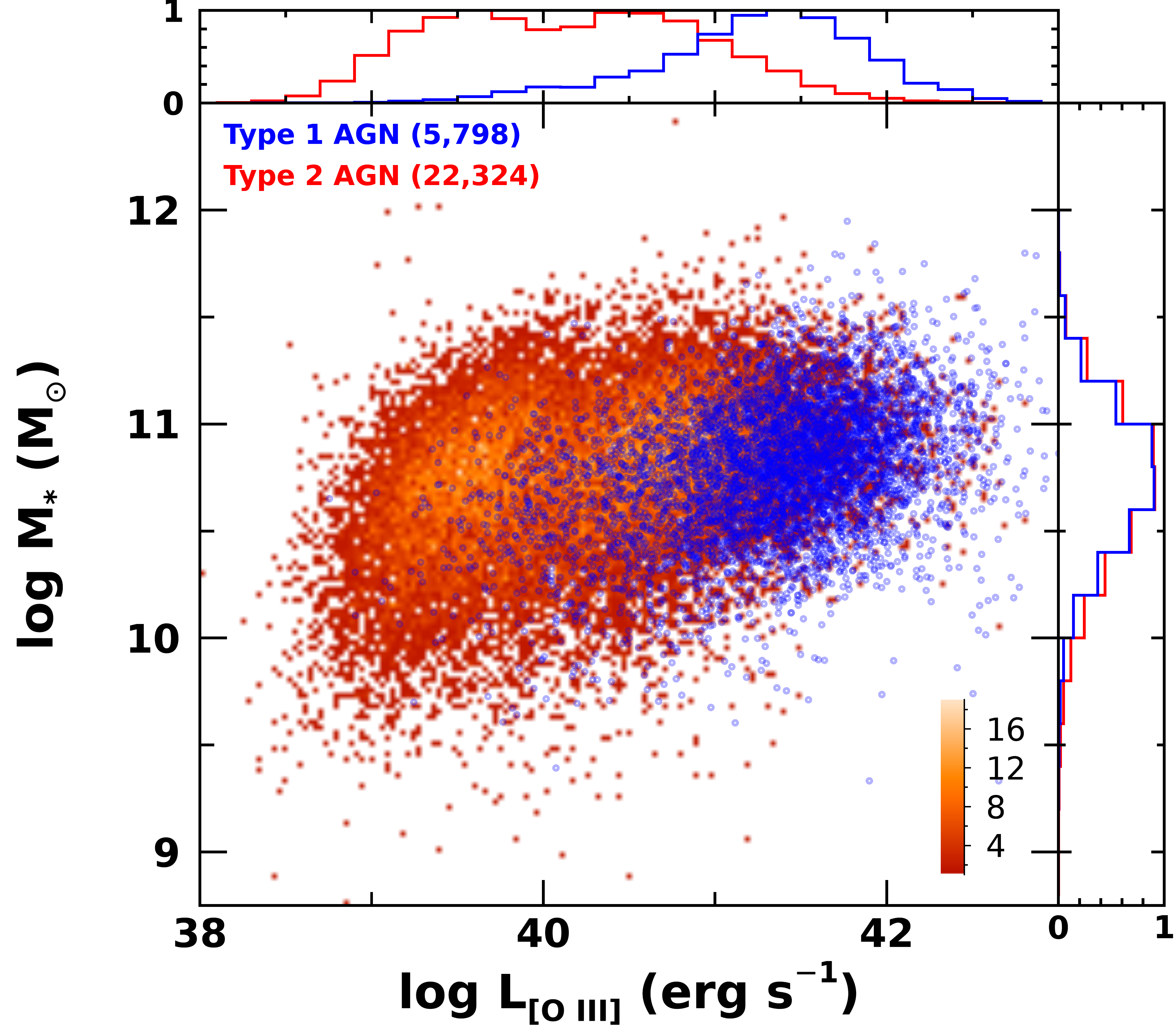}
\includegraphics[width=0.42\textwidth]{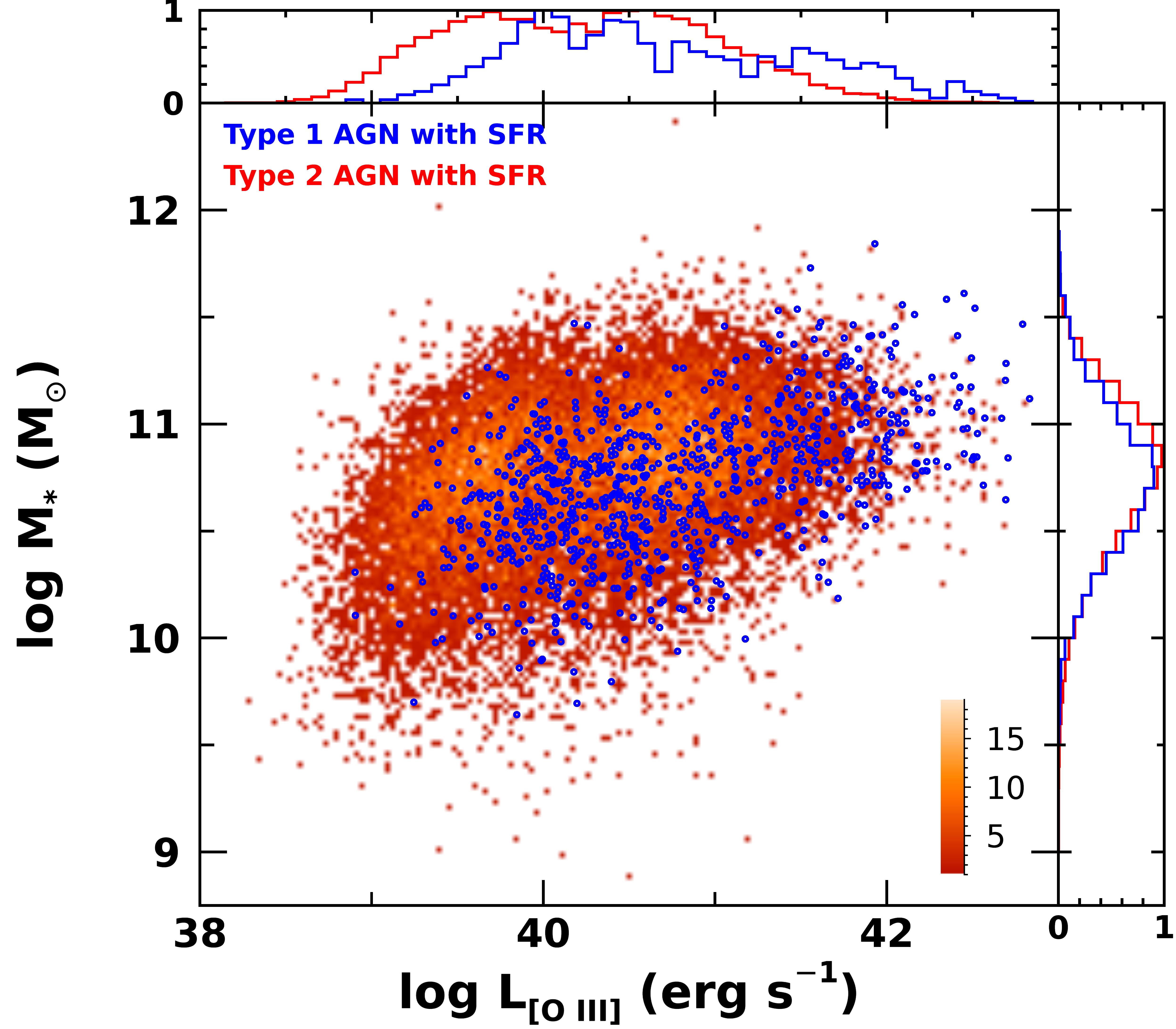}
\caption{Distributions of stellar mass and \OIII\ luminosity of type 1 (blue) and type 2 AGNs (red) for the total sample (top) and the subsample
with available SFR (bottom).
}
\end{figure}

\section{Results}

\subsection{Correlation between AGN outflows and SF}

In this section we investigate how AGN outflows are related to SF by comparing \OIII\ kinematics with SFR.
To represent the strength of \OIII\ outflows, we use the velocity dispersion (\SOIII) or velocity shift (\VOIII) measured from the \OIII\ emission line. Note that the flux-weighted \OIII\ line in the SDSS spectra is either blueshifted or redshifted, depending on the orientation of outflows and the dusty stellar disk with respect to the line-of-sight \citep[see Figure 7 of][]{Rakshit+18}. For this study, we take the absolute value of \VOIII\ to quantify the strength of outflows without considering the direction of outflows.

For this comparison we use all AGNs with available SFR, either from the direct FIR detections or based on
the ANN analysis from \citet{Ellison+16a}. 
In Figure 3 we detect a broad correlation between SFR and outflow velocities for both type 1 and type 2 AGNs,
suggesting that AGNs with stronger outflows are hosted by galaxies with higher SFR. The large scatter may indicate
that the derived SFR is relatively uncertain and/or that there are additional factors, i.e., gas fraction, Eddington ratio of AGN, local environment, etc, which change either SFR or outflow strength. 
We note that \SOIII\ shows a clearer trend than \VOIII\ as expected since velocity shift is intrinsically weak due to the cancelation between approaching and receding gas in a biconical distribution of outflows with respect to the line-of-sight. Also, the systemic velocity of type 1 AGNs suffers significant uncertainty since it is determined from the peak of \Hb\ when stellar absorption lines are not detected.

\begin{figure} 
\centering
\includegraphics[width=0.4\textwidth]{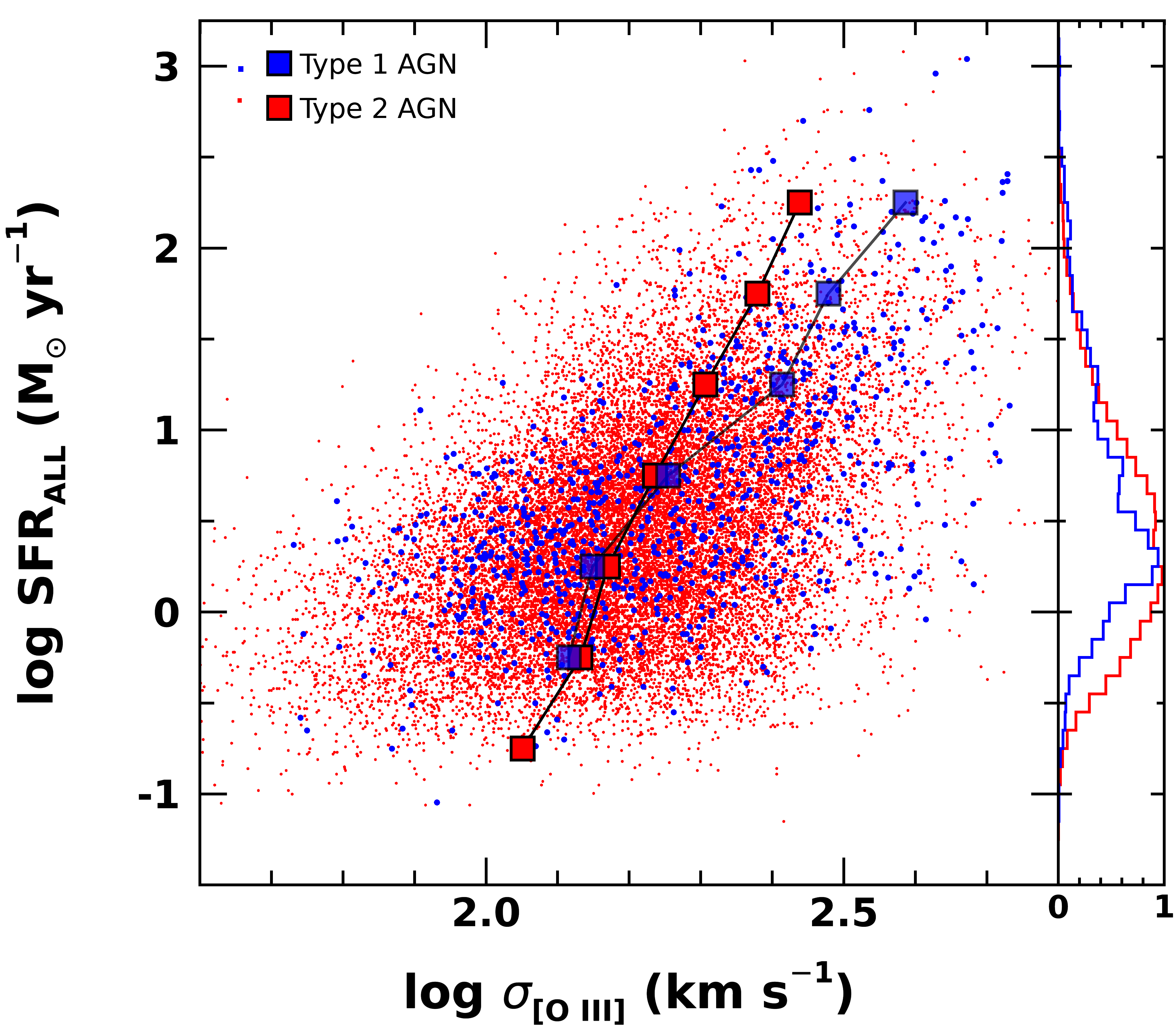}
\includegraphics[width=0.4\textwidth]{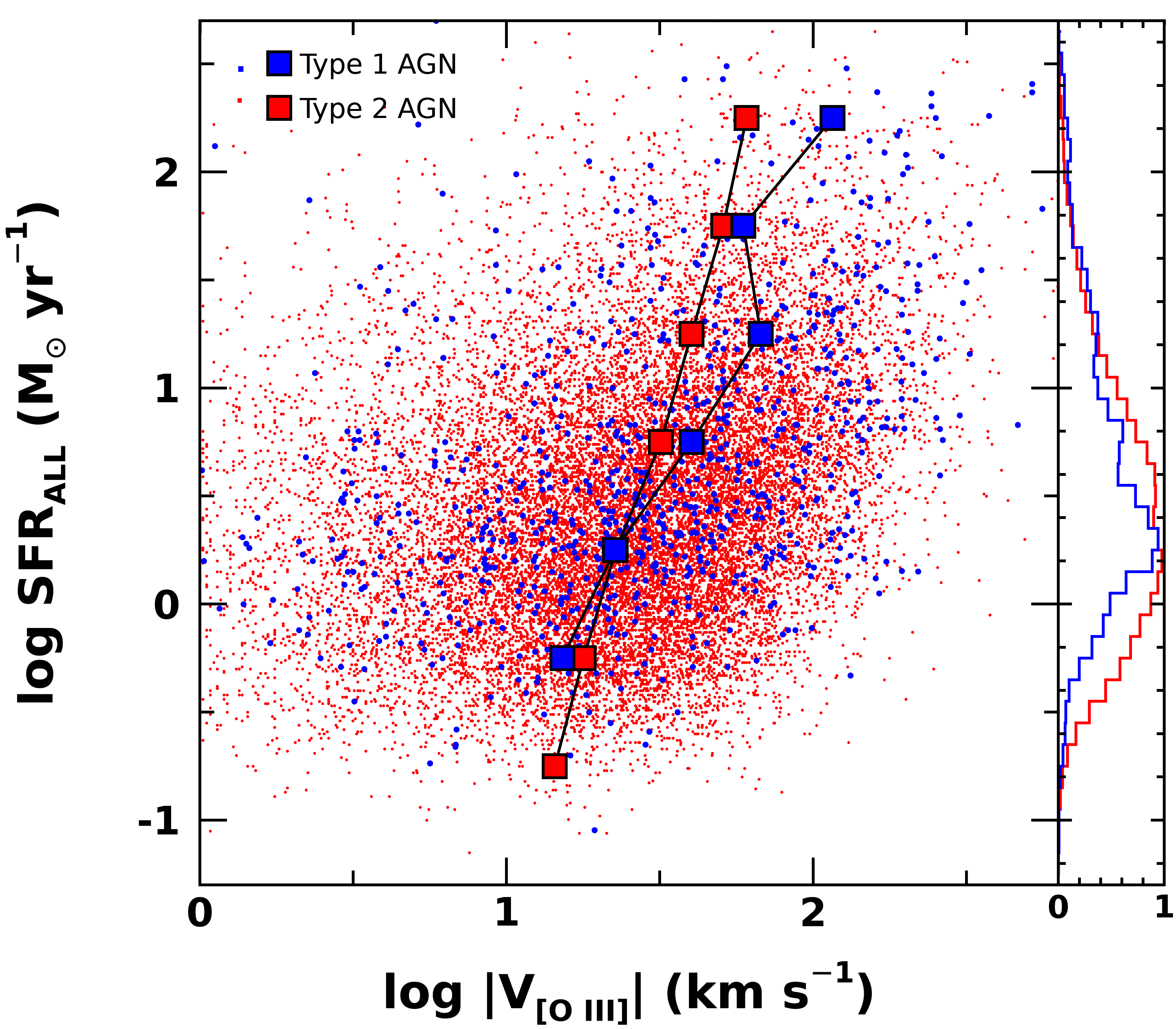}
\caption{Comparison of SFR with the velocity dispersion (top) and the absolute value of the velocity shift measured from \OIII\ (bottom) for type 1 (blue) and type 2 AGNs (red). We collect all SFR based on AKRI and Herschel FIR detection or from the ANN analysis. Large symbols are the median value of outflow velocity,
which are calculated in each bin of 0.5 dex in the y-axis. 
}
\end{figure}

\begin{figure} 
\centering
\includegraphics[width=0.45\textwidth]{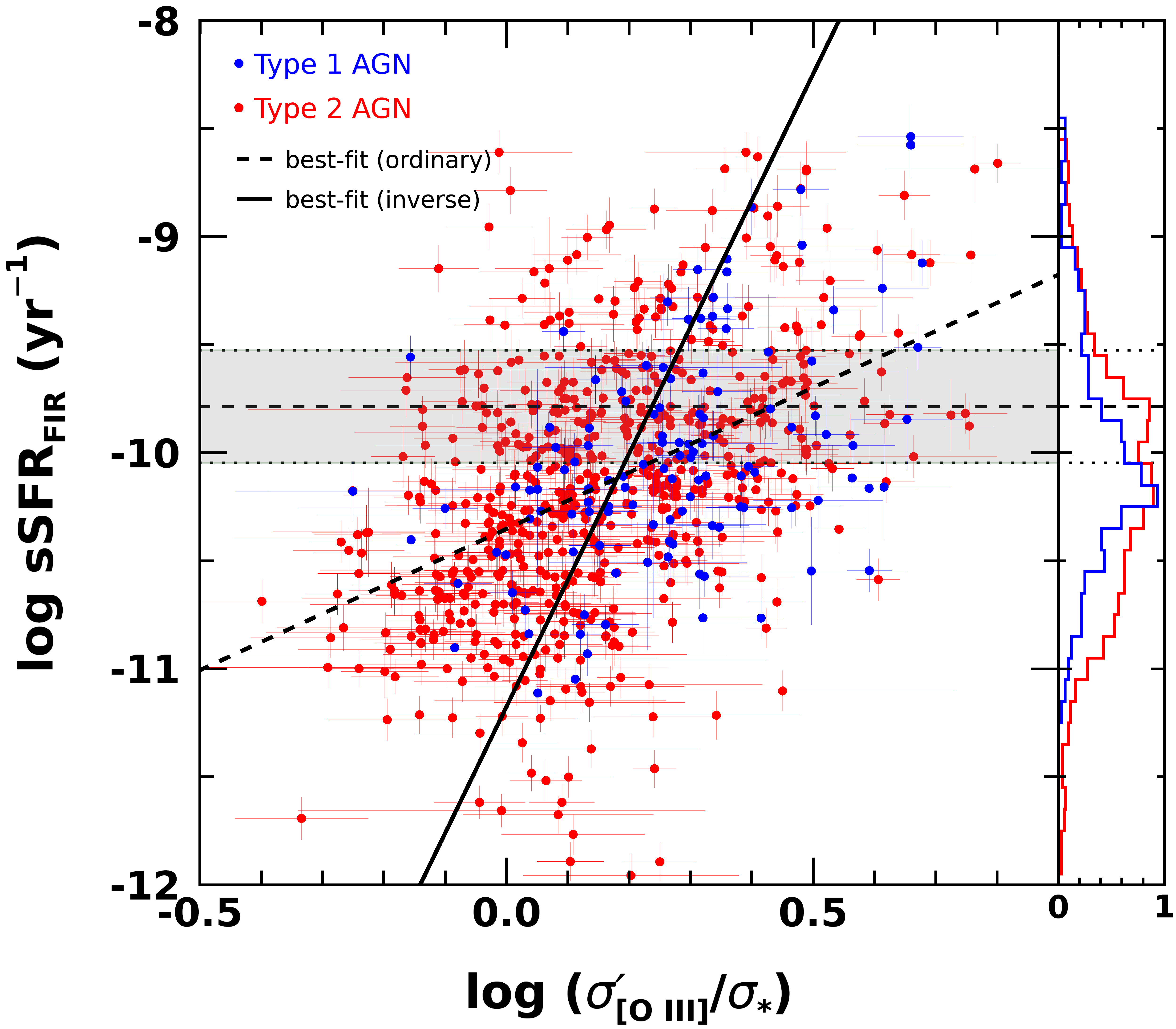}
\caption{Comparison of sSFR with outflow strength for type 1 (blue) and type 2 (red) AGNs, using only FIR based SFR. 
The best-fit slope is represented by a dashed and solid line, respectively, from the ordinary and inverse regression analysis.
The mean sSFR with 1-$\sigma$ dispersion of non-AGN SFGs is denoted with a horizontal dashed line and a grey region. 
}
\end{figure}

As \citet{Woo+16} and \citet{Rakshit+18} pointed out, the \OIII\ line profile is influenced by the gravitational potential of host galaxies. 
On average \OIII\ is systematically broader in more massive galaxies due to the rotational broadening.
Thus, we investigate how SFR correlates with outflow strength after removing the effect of gravitational potential. 
For this test, we use sSFR and the normalized \SOIII\ by stellar velocity dispersion ($\sigma_*$), in order to quantify 
the relative strength of outflows compared to the gravitational potential of host galaxies. While we are able to measure $\sigma_*$ for type 2 AGNs, $\sigma_*$ is difficult to measure for luminous type 1 AGNs owing to the strong AGN continuum. Thus, we use stellar mass as a proxy for $\sigma_*$. To avoid systematic uncertainties, we calculate $\sigma_*$ from stellar mass for both type 1 and type 2 AGNs using the best-fit relation between $\sigma_*$ and stellar mass, which is determined based on our AGN sample with measured $\sigma_*$ \citep[see Figure 2 of][]{Woo+16}. 
While we do not claim the estimated stellar velocity dispersions are accurate for individual objects, these estimates are
useful for investigating a general trend over a large dynamic range. We also note that the results remain the same when we use 
directly measured $\sigma_*$ instead of stellar mass proxies for type 2 AGNs \citep[see also][]{Woo+17}.

First, we start with a subsample of AGNs with direct FIR detections in Figure 4. For \OIII\ outflow strength,
we add velocity dispersion (\SOIII) and velocity shift (\VOIII) in quadrature to represent the outflow velocity as expressed
as $\sigma^{\prime}_{\rm OIII}$. 
For AGNs with strong outflows, \OIII\ velocity
dispersion is up to a factor of $\sim$6 higher than stellar velocity dispersion. 
We find a clear trend of higher sSFR with increasing \OIII\ outflow strength. 
We perform a regression analysis to quantify the correlation, using the \texttt{FITEXY} routine, which minimizes the $\chi^2$ statistics and accounts for measurement errors and intrinsic scatter, as well as the maximum likelihood method, which is implemented as \texttt{AMOEBA} in \texttt{IDL} \citep{1992nrfa.book.....P}. For more details, readers are advised to check the study by \citet[][see Section 2]{Park+12}, where various regression methods were compared in detail. 
Since there is a relatively large scatter and considerable errors of individual measurements, the ordinary and inverse regression results provide a different slope. The best-fit indicates that
sSFR is proportional to normalized \SOIII\ with a $5.56\pm0.31$ ($5.88\pm0.35$) or $1.54\pm0.09$ ($1.31\pm0.08$) power, respectively for inverse and ordinary regression based on the \texttt{FITEXY} (maximum likelihood) method. While the exact form of the correlation is somewhat uncertain, it is clear that sSFR is broadly correlating with outflow velocity. To compare with non-AGN galaxies, we present the mean sSFR and 1-$\sigma$ dispersion with a horizontal dashed line and a grey region, which are determined from a sample of $\sim$69,000 SFGs by \citet{Woo+17}. AGNs show a much broader distribution of sSFR, with a dependence on outflow strength.

\begin{figure*} 
\centering
\includegraphics[width=0.45\textwidth]{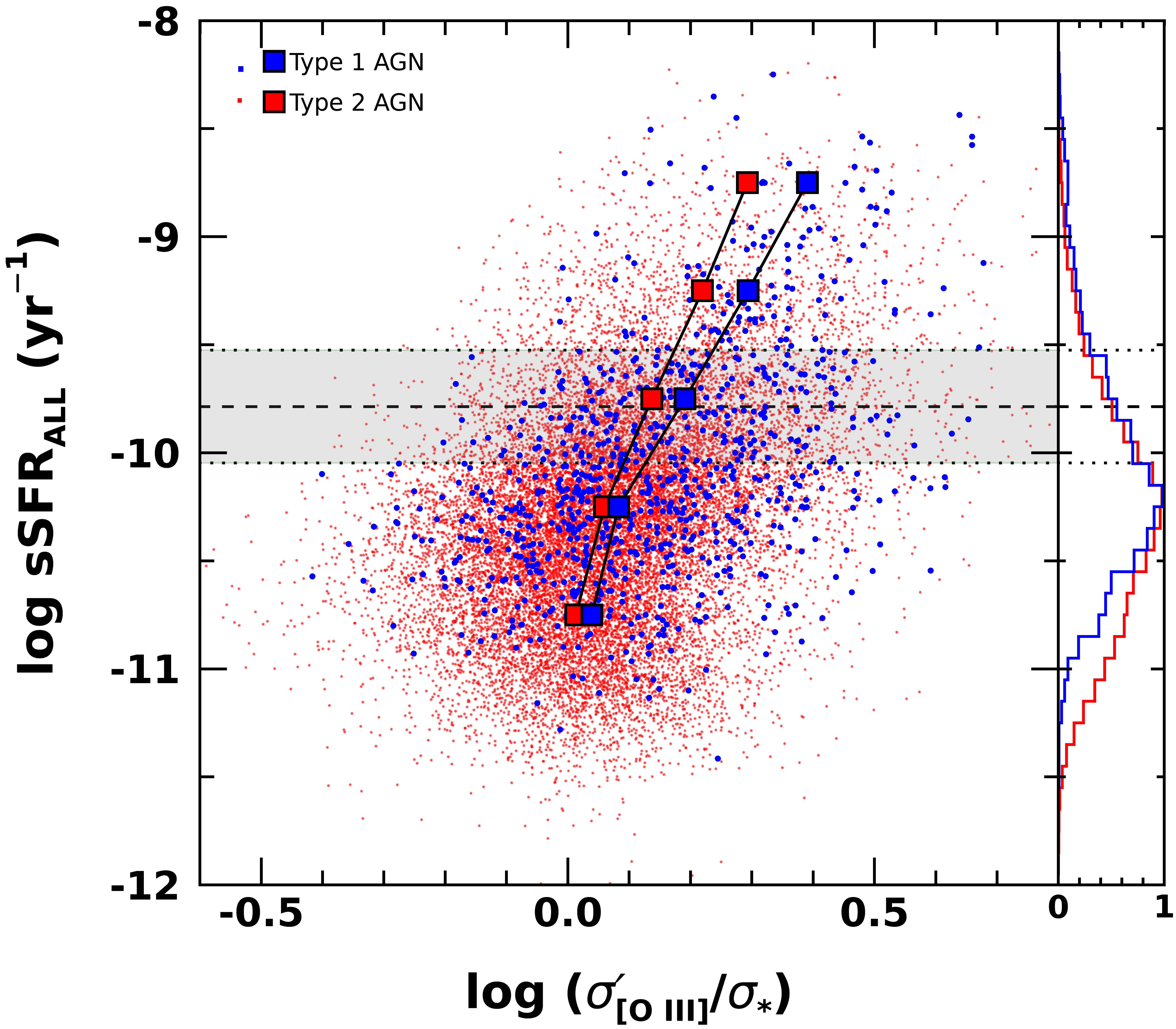}
\includegraphics[width=0.45\textwidth]{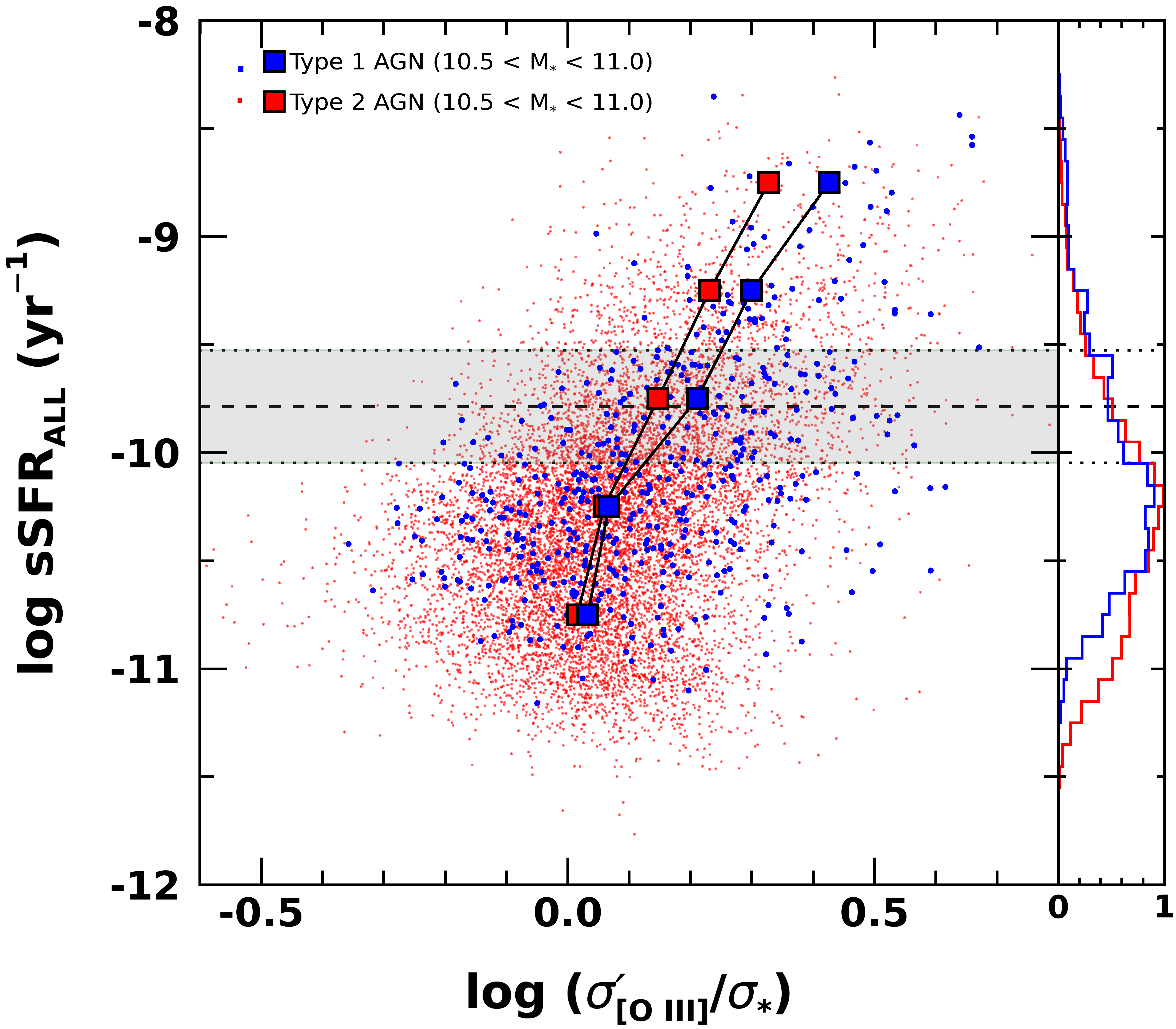}\\
\caption{
Comparison of sSFR with the normalized outflow velocity dispersion for AGNs with all available SFR (left)
and for AGNs in a limited stellar mass range (right).
Large symbols are the median value of outflow velocity,
which are calculated in each bin of 0.5 dex in the y-axis.
The mean sSFR with 1-$\sigma$ dispersion of SFGs is denoted with a horizontal dashed line and a grey region. 
}
\end{figure*}

\begin{figure} 
\centering
\includegraphics[width=0.45\textwidth]{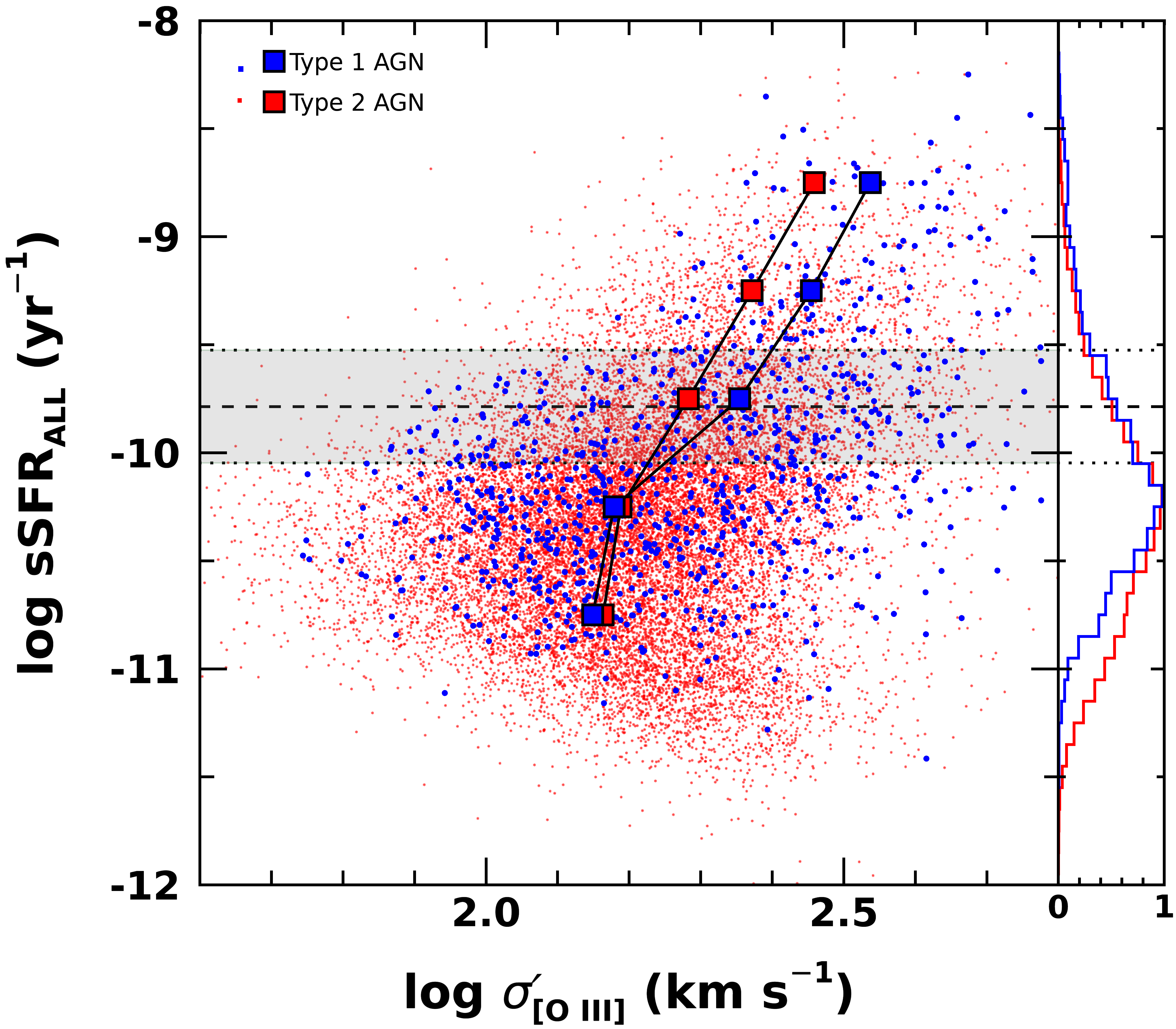}
\caption{Comparison of sSFR with the outflow velocity dispersion for AGNs with all available SFR. 
Large symbols are the median value of outflow velocity,
which are calculated in each bin of 0.5 dex in the y-axis. }
The mean sSFR with 1-$\sigma$ dispersion of SFGs is denoted with a horizontal dashed line and a grey region. 
\end{figure}

\begin{figure} 
\centering
\includegraphics[width=0.45\textwidth]{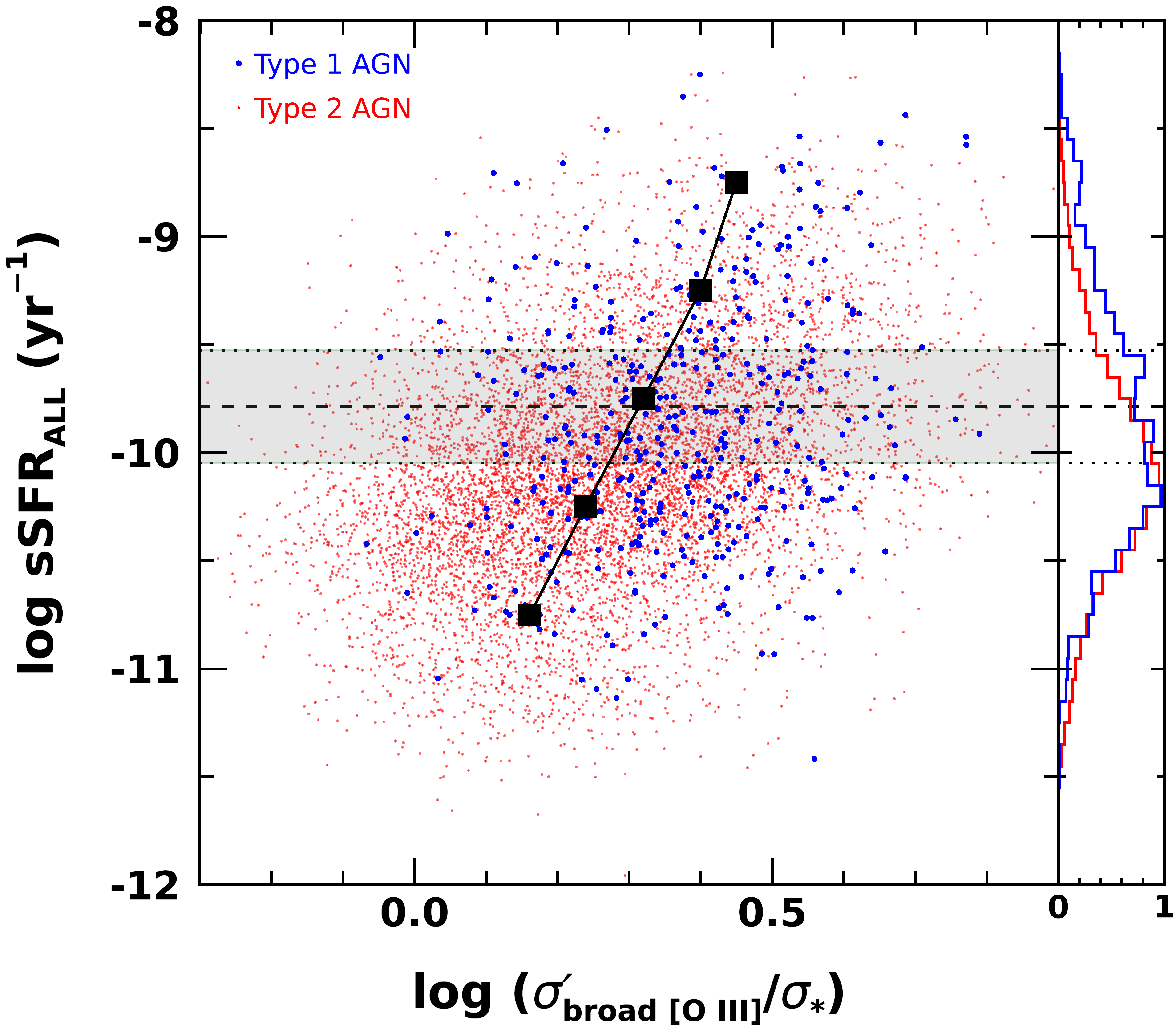}
\caption{Comparison of sSFR with the normalized outflow velocity dispersion for a subsample of AGNs with a strong broad component in \OIII.
Large symbols are the median value of outflow velocity,
which are calculated in each 0.5 dex bin in y-axis. 
Note that AGNs with \OIII, which is fitted with a double Gaussian model, are included, and the broad Gaussian component is
used for calculating gas velocity dispersion. 
Large symbols are the median value of outflow velocity,
which are calculated in each bin of 0.5 dex in the y-axis. }
\end{figure}

Second, we enlarge the sample by including AGNs without FIR detections, but with SFR estimates from the ANN
analysis \citep{Ellison+16a}. In the left panel of Figure 5, we show $\sim$20,000 AGNs with the estimated SFR with $\sigma_{ANN}$$<$ 0.3. Albeit with a large scatter, we find a positive correlation that stronger outflow AGNs 
are hosted by galaxies with higher SFRs. As a function of sSFR, we calculate the median value of the normalized \OIII\ velocity dispersion, which are denoted with large squares. A strong trend represented by the median values indicates that sSFR is higher for AGNs with stronger outflows, except for the galaxies with very low sSFR (e.g., sSFR $<$ 10$^{-11}$). Note that these low SF galaxies are mostly early-type galaxies \citep[see][]{Woo+17}. 
To check whether the trend is driven by galaxy mass, we limited the sample by selecting targets within 10.5 $<$ log M$_{*}$ $<$ 11 in the right panel of Figure 5. 
For this subsample, we find virtually the same correlation.

Compared to SFGs (a grey region in Figure 5), AGNs show a much broader distribution of sSFR,
and the mean sSFR is lower than that of SFGs. However, depending on the outflow strength, there is an interesting trend. AGNs with strongest outflows (i.e., highest \OIII\ velocity dispersion) are located above the main sequence, while AGNs with weaker/no outflows (i.e., lowest \OIII\ velocity dispersion) show lower sSFR than main sequence SFGs. Our previous study reported the same trend for type 2 AGNs \citep{Woo+17}, and here we confirm the correlation for the combined sample of type 1 and type 2 AGNs.
The positive correlation between outflow strength and SFR is inconsistent with a negative feedback scenario, which may predict an opposite correlation that AGNs with stronger outflows have much lower sSFR than that of SFGs. On the other hand, AGN with weak/no outflows are show lower sSFR than main sequence SFGs. This result may suggest that AGN feedback is not instantaneous but the suppression of star formation is observable only after a certain feedback time scale, which may depend on the detailed mechanism of suppression, the size and location of the SF regions in host galaxies (i.e., circumnuclear vs. disk), etc. 

As a consistency check, we investigate whether the correlation of sSFR with the normalized \SOIII\ is artificially driven by the large dynamic range of stellar mass.
Since both SFR and \SOIII\ are normalized by stellar mass, a broad trend may be introduced even if SFR and \SOIII\ are not intrinsically correlated. 
This is not the case since we already presented the correlation between SFR and \SOIII\ in Figure 3. Nevertheless, we
investigate the possibility by using \SOIII, instead of the normalized \SOIII\ in Figure 6. We find a very similar distribution 
and a consistent trend, indicating that the correlation between sSFR and outflows is valid. 
Note that most early-type galaxies with very low sSFR show relatively large \SOIII, since the contribution from the gravitational broadening is significant 
due to the relatively high galaxy mass. 
When we limit the stellar mass range as 10.5 $<$ log M$_*$ $<$ 11, we obtain qualitatively the same results as in Figure 6.

Third, for comparing with the results in the literature, we use a broad component of \OIII\ after removing a narrow component by 
assuming that the narrow component represents the gravitational potential of host galaxies.
For this experiment, we use a subsample of AGNs, for which \OIII\ is fitted with a double Gaussian model. Among these AGNs, the two components of \OIII\ show a similar width for some case, leading to difficulty to separate gravitational and non-gravitational components. Thus, we only select when the broad component of \OIII\ is broader by more than a factor of two
than the narrow component. At the same time, we only consider when the broad component of \OIII\ is fairly broad, i.e., \SOIII\ $>$ 100 \kms. With these criteria, we selected a subsample of 8,171 type 2 AGNs and 435 type 1 AGNs, 
in order to compare sSFR with the outflow strength measured from the broad wing component of \OIII\  (Figure 7). 

As expected the outflow strength is much higher when the broad component is used in calculating  gas velocity dispersion. 
The mean log ratio of gas-to-stellar velocity dispersion is $0.27\pm0.20$, suggesting that outflow gas velocity dispersion is
typically a factor of two higher than stellar velocity dispersion for this subsample. We clearly see an increasing trend of sSFR with increasing outflow
strength. When the gas-to-stellar velocity dispersion ratio is higher than two, sSFR is higher than that of SFGs.   
In contrast, for weaker outflow AGNs, sSFR is lower than that of SFGs. We obtained the best-fit relation with a $7.14\pm0.20$ slope
based on the \texttt{FITEXY} inverse regression, while the ordinary regression provides the best-fit with a slope of $1.00\pm0.03$. 
Note that stellar velocity dispersion is estimated based on stellar mass, hence, the systematic uncertainty of $\sigma_{*}$ is partly responsible 
for a large scatter (i.e., 0.25 dex in the velocity dispersion ratio). Nevertheless, there is a clear positive correlation between sSFR and outflow
strength. We divide the subsample into five bins based on the sSFR, and calculate median gas-to-stellar velocity dispersion ratio
as shown in Figure 7. For the lowest bin (i.e., log sSFR = -10.75), the mean log velocity dispersion ratio is $0.17\pm0.17$ (i.e., a factor of $\sim$1.5)
while for the highest bin (i.e., log sSFR = -8.75), the mean velocity dispersion ratio is $0.45\pm0.19$ (i.e., a factor of $\sim$2.8).

\subsection{sSFR dependence on Eddington ratio and outflow strength}

In this section we investigate how the sSFR systematically changes depending on AGN energetics and outflow strength. 
To determine the Eddington, we calculate bolometric luminosity using AGN continuum luminosity at 5100\AA\ with a bolometric correction factor
10 for type 1 AGNs \citep{Woo+02}, and the \OIII\ luminosity as a proxy for bolometric luminosity of type 2 AGNs \citep[for more details see][]{Woo+17}.

First, we divide AGNs into four Eddington ratio bins in order to explore the effect of AGN energetics in comparing with SFGs. 
In Figure 8, we present the distribution of sSFR of AGNs with respect to that of SFG. The distribution of SFGs is determined using a sample of $\sim$69,000 non-AGN emission line galaxies, which were identified based on the emission line ratio diagrams by \citet{Woo+17}. Here, we show the relative difference of sSFR ($\Delta$sSFR) with respect to the mean sSFR of SFGs (black vertical line). We find a clear systematic trend that the sSFR decreases with decreasing Eddington ratio for both type 1 and type 2 AGNs. For high Eddington ratio AGNs (i.e., 1\% -10\%), the distribution of sSFR is comparable to SFGs, while AGNs with higher Eddington ratio (i.e., $>$ 10\%)
show slightly higher sSFR than SFGs, albeit with a broad dispersion. In contrast, for lower Eddington ratio AGNs, the sSFR distribution
clearly shifts to lower values, presenting a lower mean sSFR by more than 0.5 dex than SFGs. For AGNs with the lowest Eddington ratios (i.e., $<$0.1\%), the mean sSFR is smaller by more than a factor of ten. Note that Eddington ratio is calculated differently for type 1 and type 2 AGNs.
Nevertheless, we generally find similar distributions at each bin
and a consistent trend of decreasing sSFR with decreasing Eddington ratio for both type 1 and type 2 AGNs. 

\begin{figure} 
\centering
\includegraphics[width=0.42\textwidth]{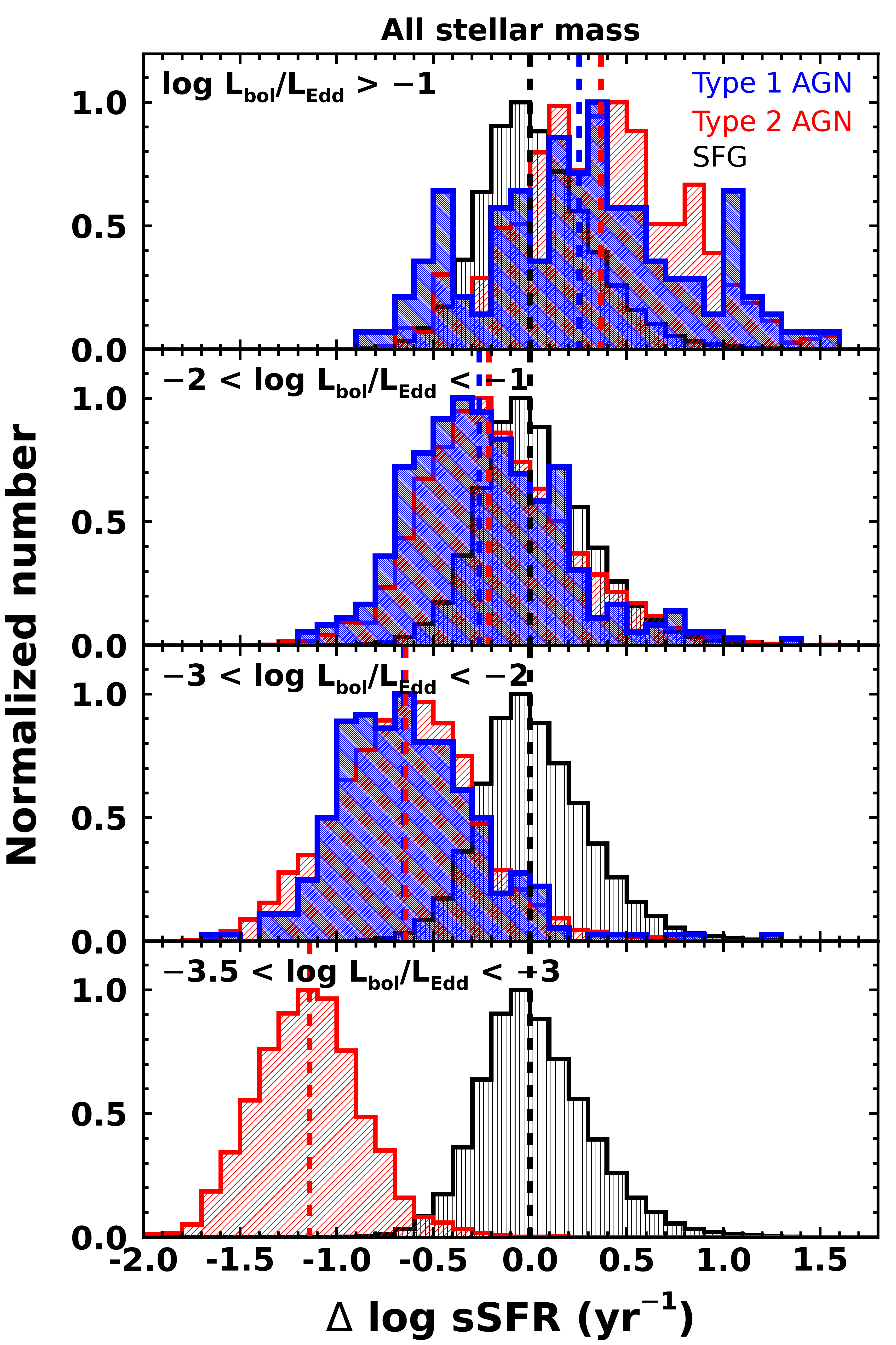}

\caption{Distribution of the relative sSFR depending on Eddington ratio of type 1 (blue) and type 2 AGNs (red), with respect to SFGs (gray).
The mean value of each distribution is denoted with vertical dashed lines.  
High Eddington AGNs ($>$0.1) tend to show higher sSFR than SFGs, suggesting enhanced SF, while low Eddington AGNs
clearly show much lower sSFR than SFG.
}
\end{figure}  

\begin{figure*} 
\centering
\includegraphics[width=0.45\textwidth]{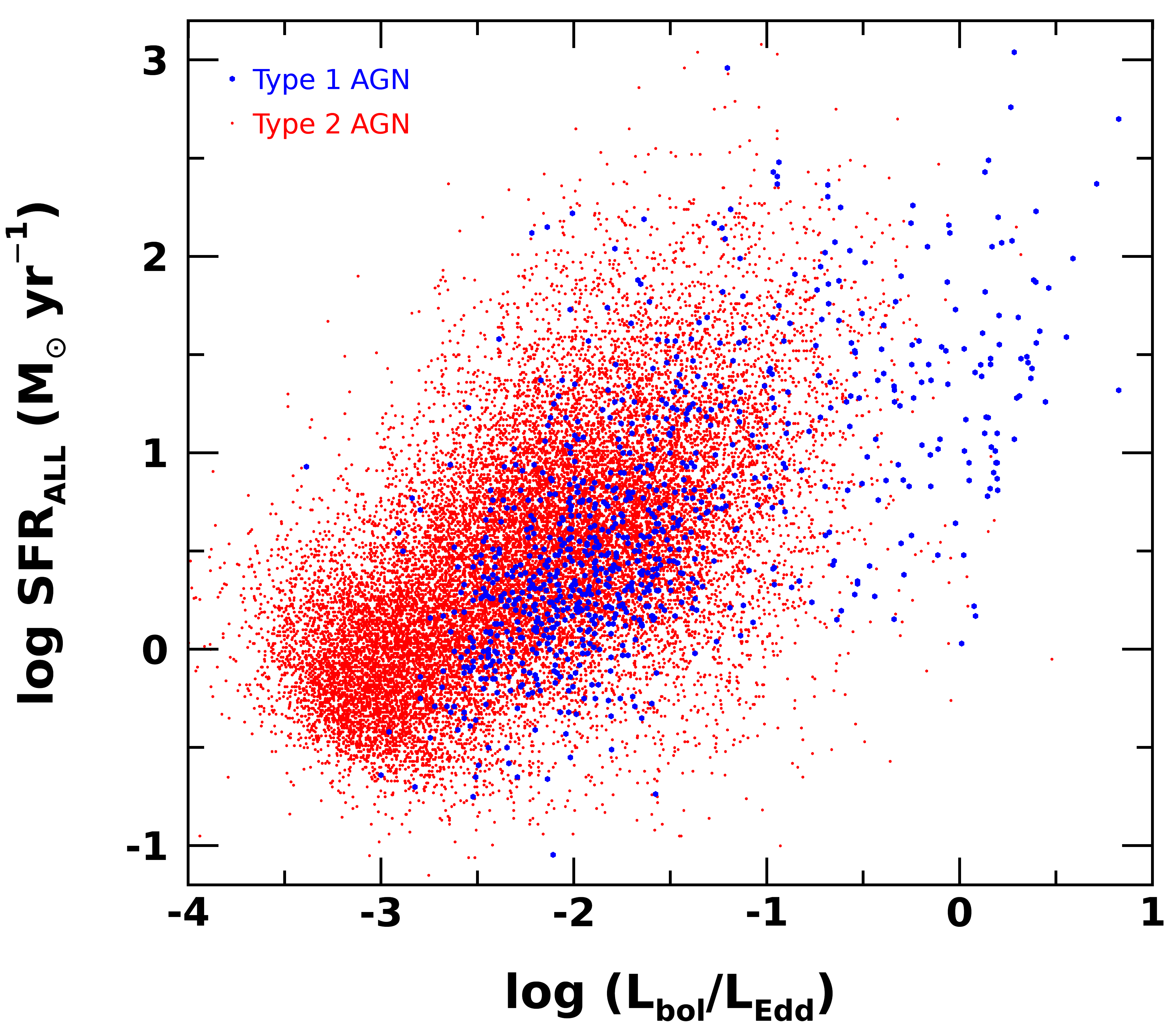}
\includegraphics[width=0.45\textwidth]{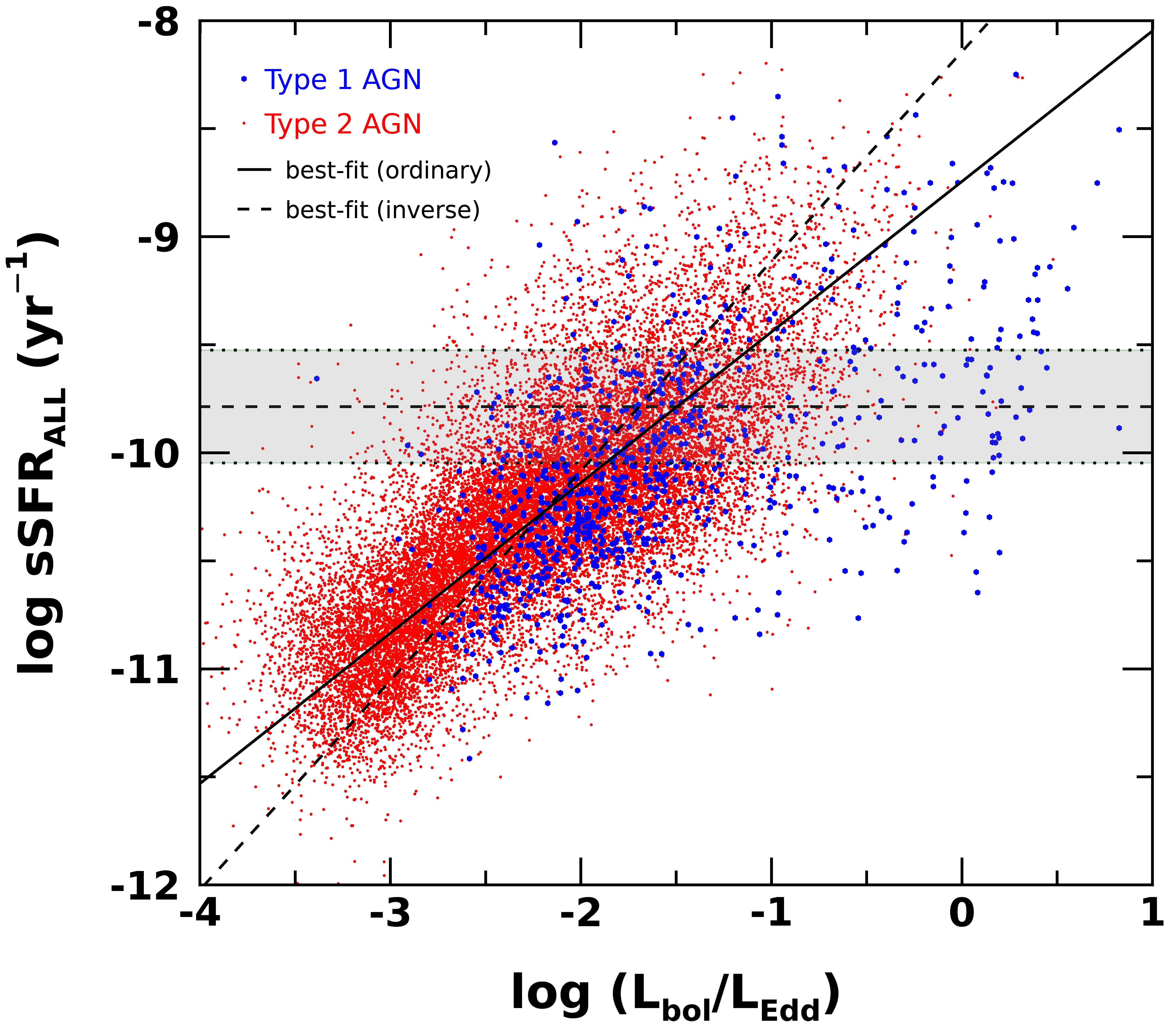}
\caption{Comparison of Eddington ratio with SFR (left) or sSFR (right) for type 1 (blue) and type 2 AGNs (red). 
For the sSFR vs. Eddington ratio comparison, the best-fit is represented by a thick line (ordinary regression) or a thin line (inverse regression). 
We used a horizontal line and a grey region to denote the mean sSFR of non-AGN galaxies and their 1 $\sigma$ dispersion. 
}
\end{figure*}

In Figure 9 we directly compare Eddington ratio with either SFR or sSFR. 
There is a broad correlation of increasing SFR with increasing Eddington ratio albeit with a large scatter. 
From the ordinary (inverse) regression we obtain that SFR is proportional to Eddington ratio with a 0.33$\pm$0.02 (1.47$\pm$0.01) power
and a total scatter of 0.53 (0.52) dex and an intrinsic scatter of 0.50$\pm$0.01 (0.43$\pm$0.02) dex in the SFR dimension,
based on the maximum likelihood estimator which we used in \S~3.1.
For this regression analysis, we assume an average error of 0.4 dex for Eddington ratio adopting from the systematic uncertainty of
single-epoch black hole mass \citep{2012ApJ...747...30P} since the measurement uncertainties of the black hole mass and bolometric luminosity are difficult to quantify and the systematic uncertainty
of the black hole mass is dominant in determining Eddington ratio. 
If we perform the regression without using errors, we obtain virtually the same result. 

In the case of sSFR, we find a tighter correlation with Eddington ratio (left panel in Figure 9). 
The best-fit slop is 0.69$\pm$0.02 with a total scatter of 0.38 dex and an intrinsic scatter of 0.24$\pm$0.01dex from the ordinary regression,
while the inverse regression provides a slightly higher slope of 0.97$\pm$0.01 with similar scatters. 
When we restrict the sample with more reliable SFR based on either direct FIR detections or ANN-based SFR with the best quality (i.e., $\sigma_{ANN}$ $<$ 0.1),
we obtain almost the same best-fit relation (0.72$\pm$0.02 or 0.95$\pm$0.01, respectively for the ordinary and inverse regression) and
a smaller intrinsic scatter ($\sim$0.23 dex). 
These results indicate that SFR per stellar mass is closely linked to AGN luminosity per black hole mass.

Second, we divide AGNs into 3 bins depending on the strength of outflows in order to investigate how the distribution of sSFR changes
with outflow strength. As a tracer of outflow strength, we use \OIII\ velocity dispersion normalized by stellar velocity dispersion, as we used in the previous sections. 
We define strong (i.e., $\sigma_{\rm [OIII]}'$/\SVD $>$2), intermediate (i.e., 1$<$ $\sigma_{\rm [OIII]}'$/\SVD$<$ 2), and weak outflows (i.e., $\sigma_{\rm [OIII]}'$/\SVD $<$1) in Figure 9. Similar to the effect of Eddington ratios, we find a clear decreasing trend of sSFR with decreasing outflow strength. For both type 1 and type 2 AGNs, strong outflow AGNs
show comparable sSFR with respect to SFGs, albeit with a large dispersion (top panel of Figure 9). If we select strongest outflow AGNs with $\sigma_{\rm [OIII]}'$/\SVD $>$3, we obtain the mean $\Delta$sSFR of $0.42\pm0.62$ and  $0.10\pm0.50$, respectively for type 1 and type 2 AGNs, suggesting that SF is enhanced 
by a factor of 2-3 compared to non-AGN galaxies. 
For AGNs with intermediate outflows, we find somewhat lower mean sSFR than SFGs (see also Table 2). 
The difference of the mean sSFR with respect to SFGs systematically increases for weaker outflow AGNs (bottom panel of Figure 9). The slightly different mean sSFR between type 1 and type 2 AGNs is presumably caused by the sample selection. Since the type 1 AGN sample has on average higher Eddington ratios, sSFR is also somewhat higher for a given subsample. Note that the sSFR distribution of type 2 AGNs extends to much lower value than that of type 1 AGNs due to the sample selection.

Since the stellar mass distributions of the AGN and SFG samples are different, we test whether the different galaxy mass scale affect the trend, by using a subsample of AGNs in a limited stellar mass  bin (10.5$<$ log M$_{*}$/M$_{\odot}$ $<$ 11). We find virtually the same result of decreasing sSFR with decreasing Eddington ratio as well as decresing outflow strength \citep[see also Figure 9 and 10 of][]{Woo+16}.  
Thus, we conclude that there is an intrinsic trend of decreasing sSFR for AGNs with weaker outflows.

\begin{table}
\begin{center}
\caption{mean $\Delta$sSFR depending on Eddington ratios in Fig. 7}
\begin{tabular}{lrr}
\tableline
\tableline
log L$_{\rm bol}$/L$_{\rm Edd}$				&	\multicolumn{2}{c}{$\Delta$ log sSFR (yr$^{-1}$)}	\\
										&	Type 1				&	Type 2	\\
\tableline
log L$_{\rm bol}$/L$_{\rm Edd}$ $ > -1$			&	$0.25 \pm 0.53$		&	$0.37 \pm 0.43$	\\
$-2 <$ log L$_{\rm bol}$/L$_{\rm Edd}$ $< -1$		&	$-0.26 \pm 0.40$	&	$-0.21 \pm 0.39$	\\
$-3 <$ log L$_{\rm bol}$/L$_{\rm Edd}$ $< -2$		&	$-0.65 \pm 0.38$	&	$-0.64 \pm 0.37$	\\
$-3.5 <$ log L$_{\rm bol}$/L$_{\rm Edd}$ $< -3$	&	-				&	$-1.14 \pm 0.28$	\\
\tableline
\end{tabular}
\end{center}
\tablecomments{
}
\end{table}

\begin{table}
\begin{center}
\caption{mean $\Delta$sSFR depending on outflow strengths in Fig. 8}
\begin{tabular}{lrr}
\tableline
\tableline
log ($\sigma^\prime_{\rm [O III]}$/$\sigma_{*}$)				&	\multicolumn{2}{c}{$\Delta$ log sSFR (yr$^{-1}$)}	\\
													&	Type 1			&	Type 2	\\
\tableline
log ($\sigma^\prime_{\rm [O III]}$/$\sigma_{*}$) $> 0.5$			&	$0.42 \pm 0.62$		&	$0.10 \pm 0.50$	\\
$0.3 <$ log ($\sigma^\prime_{\rm [O III]}$/$\sigma_{*}$) $< 0.5$	&	$0.08 \pm 0.56$		&	$-0.10 \pm 0.54$	\\
$0 <$ log ($\sigma^\prime_{\rm [O III]}$/$\sigma_{*}$) $< 0.3$		&	$-0.35 \pm 0.49$	&	$-0.49 \pm 0.52$	\\
log ($\sigma^\prime_{\rm [O III]}$/$\sigma_{*}$) $< 0$			&	$-0.58 \pm 0.33$	&	$-0.69 \pm 0.43$	\\
\tableline
\end{tabular}
\end{center}
\tablecomments{
}
\end{table}

\begin{figure} 
\centering
\includegraphics[width=0.42\textwidth]{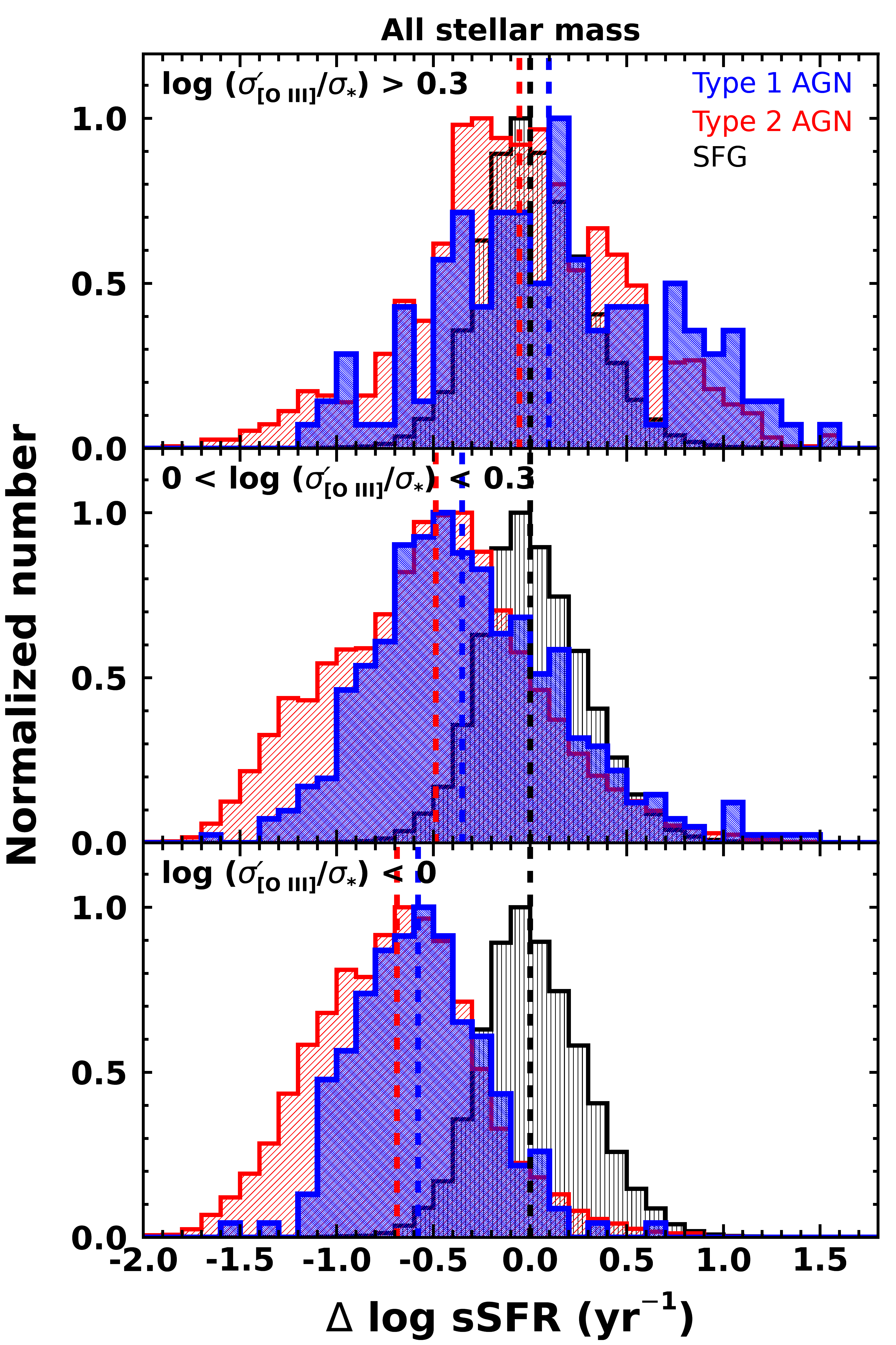}
\caption{Distribution of sSFR depending on outflow strength of type 1 (blue) and type 2 AGNs (red), with respect to SFGs (gray). 
The mean value of each distribution is denoted with vertical dashed lines.  
Strong outflow AGNs tend to have similar sSFR compared to SFGs, while no outflow AGNs show much lower sSFR.
}
\end{figure}

\section{Discussion}

\subsection{gas feeding or delayed feedback?}

We find that
AGNs with high Eddington ratio typically show strong gas outflows and at the same time these AGNs are hosted by typical SFGs in the main sequence. In contrast, AGNs with low Eddington ratio show weak or no outflows
and their host galaxies have systematically lower SFR. The apparent connection between SFR and AGN energetics was noted by previous studies 
with variously selected samples based on optical, mid-IR, radio and X-ray characteristics \citep[e.g.][]{Shimizu+15, Ellison+16b} as these samples have somewhat different 
mean Eddington ratios. 
The correlation between SFR and outflow kinematics as well as Eddington ratio was reported for a large sample of type 2 AGNs by our previous work \citep{Woo+17}. The connection between the two was interpreted as that gas outflows play a role as the mechanism of delivering the energy output from the central black hole to gas in a larger scale of host galaxies, affecting star formation activity. The positive correlation of outflow strength with SFR indicates no clear sign of instantaneous or fast feedback of suppressing SF. 
On the other hand, the systematically lower SFR of the galaxies, that host low Eddington AGNs with no outflows, can be explained by a delayed feedback scenario. When AGN energetics are strong, SFR is also high as the gas in the host galaxy feeds black hole and SF at the same time. After certain time scale, however, which is required for outflows to start impacting on gas in a large scale, both AGN accretion, hence the strength of outflows, and SFR becomes weaker.

In this study we find the same correlation between AGN outflows and SFR for type 1 and type 2 AGN samples. Since optical type 2 AGNs were selected based on the emission line ratio criteria, the sample may suffer incompleteness in low luminosity. In addition, black hole mass and bolometric luminosity are more uncertain than those of type 1 AGNs. In order to overcome these limitations, we combine type 1 and type 2 AGNs, and obtain the same relation among SFR, outflow strength, and Eddington ratio. We find a relatively tight correlation between sSFR and Eddington ratio, indicating that SFR per galaxy mass is closely linked to mass accretion rate per black hole mass. As we showed that the outflow strength, in particular \SOIII\ measured from the width of \OIII, is much higher than that of non-AGN galaxies \citep[see Fig. 6 and 7 of][]{Woo+17}, the outflows are driven by AGNs. Thus, the kinematical strength of AGN-driven outflows are correlated with SFR, depending on the Eddington ratio of AGNs. 
The correlation among accretion rate, outflow strength, and SFR may suggest that gas supply feeds both black hole and SF activity at the same time, while the decrease of gas supply or gas consumption would
weaken black hole accretion, outflows, as well as SF.
While star formation is observed in a large scale disk or within a circumnulcear region \citep[Kennicutt+98], the majority of our sample is likely to be disk galaxies with a moderate IR luminosity (SFR$<$ $\sim$100 \msun yrs$^{-1}$). The fraction of interacting or starburst galaxies is not explored in this study, while we expect that disk star formation is dominant among our galaxies at z $<$ 0.3.

The correlation between sSFR and outflow strength may support a positive feedback scenarios for higher Eddington AGNs with stronger outflows. 
In contrast, a negative feedback scenario may explain the trend that lower Eddington AGNs with weaker outflows have much lower SFR (Figure 7 and 8). Without invoking AGN feedback, It is also possible to interpret the correlation as a result of changing gas fraction. When gas fraction is high, feeding black hole and star formation is more efficient, leading to high accretion and strong outflows as well as strong SF activity. In contrast, when gas fraction is low, feeding is much less efficient, resulting in low Eddington AGNs with weak outflows and much lower SFR. Thus, the correlation may be a natural consequence of decreasing gas fraction in a given galaxy or increasing gas fraction due to external mechanisms such as accretion from neighbor galaxies or filaments in a large scale structure. It is unclear whether the decrease of gas fraction is due to the mechanical work by AGN gas outflows or due to the gas consumption for forming new stars. It is also possible that there is an intrinsic difference of gas fraction among various galaxies at fixed stellar mass. Note that these scenarios do not require AGN feedback. Thus, we conclude that while we find a clear correlation between SFR and AGN outflow kinematics, there is no strong
evidence of AGN feedback regulating SFR, at least in the local universe. 

It would be interesting to investigate the systematic difference of gas fraction along the correlation between SFR and outflow kinematics, however, 
direct constrain on the gas fraction seems difficult since available gas surveys, i.e, HI survey to probe cold gas fraction, are relatively shallow \citep[e.g.,][]{Jones+18, Catinella+18}. 
Note that global gas fraction in galaxies may not be directly related to AGN activity since the local feeding at the very center around a black hole may be more important. The local feeding is not necessarily related to large scale gas distribution, while gas fraction is clearly linked to the global SFR. Thus, gas fraction can only describe the correlation between SFR and outflow kinematics as a general trend.

\section{Summary \& Conclusion}

We investigate the correlation between SF and AGN activities using a large sample of type 1 and type 2 AGNs.
By comparing AGN outflow kinematics and energetics with SFR, we find a systematic change of sSFR of AGNs with respect to that of SFGs in the main sequence. 
We summarize the main results as follows.

\medskip

$\bullet$ Using the SFR determined from FIR detections or estimated based on the ANN analysis, we find a correlation between
SFR and either \OIII\ velocity dispersion or velocity shift, indicating that AGNs with stronger outflow strength tend to be hosted by galaxies with 
higher SFRs. 

$\bullet$ We find that sSFR correlates with the normalized gas velocity dispersion by stellar velocity dispersion.
AGNs with strongest outflows are located above the main sequence of SFGs, while
AGNs with weak or no outflows show much lower sSFR, suggesting no sign of instantaneous (or short-time scale) feedback suppressing star formation.

$\bullet$ The distribution of sSFR shows a clear trend with Eddington ratio and outflow strength. Both type 1 and type 2 AGNs with 1\% or higher Eddington ratio show comparable sSFR with respect to SFGs. In contrast, lower Eddington AGNs clearly show much lower sSFR. We find a correlation between sSFR and Eddington ratio, indicating SF per galaxy mass is linked to AGN accretion per black hole mass.

$\bullet$ Strong outflow AGNs show similar sSFR compared to SFGs in the main sequence. In contrast, no outflow AGNs have much lower sSFR than SFGs.
These results imply that AGN feedback is delayed or gas feeding to both AGN and SF is limited due to gas consumption.

\acknowledgments
We thank the anonymous referee for detailed comments, which were useful to improve the clarity of the manuscript. 
Support for this work was provided by the National Research Foundation of Korea grant funded by the Korea government (No. 2016R1A2B3011457).

\bibliographystyle{apj}

\end{document}